\documentclass[letterpaper, 12pt]{article}
\usepackage{jheppub}

\usepackage{graphicx}
\usepackage{bbm}
\usepackage{float}
\usepackage{amssymb}
\usepackage{amsthm}
\usepackage{multirow}
\usepackage{physics}
\usepackage{afterpage}
\usepackage{rotating}
\usepackage{upgreek}
\usepackage{amsmath}
\usepackage{mathtools}
\usepackage{epstopdf}
\usepackage{mathrsfs}
\usepackage[labelformat=simple]{subcaption}
\usepackage{calligra}
\usepackage[T1]{fontenc}  

\begin{document}

\begin{titlepage}

\setcounter{page}{1} \baselineskip=15.5pt \thispagestyle{empty}
{\flushright {ITP-CAS-25-292}\\}
		
\bigskip\
		
\vspace{1.4cm}
\begin{center}
{\LARGE \bfseries Some Properties of Multi-Component Axion\vspace{0.24cm}\\ Dark Matter}
\end{center}
\vspace{0.15cm}
			
\begin{center}
{\fontsize{14}{30}\selectfont Hai-Jun Li$^{a,b}$}
\end{center}
\begin{center}
\vspace{0.25 cm}
\textsl{$^a$Institute of Theoretical Physics, Chinese Academy of Sciences, Beijing 100190, China}\\
\textsl{$^b$International Centre for Theoretical Physics Asia-Pacific, Beijing 100190, China}\\

\vspace{-0.1 cm}				
\begin{center}
{E-mail: \textcolor{blue}{\tt {lihaijun@itp.ac.cn}}}
\end{center}	
\end{center}
\vspace{0.6cm}
\noindent

We introduce a mechanism for multi-component dark matter (DM) that originates from axion mixing and present some of its defining properties.
In this context, multi-component DM implies that the cold DM is composed of the QCD axion and many ultra-light axion-like particles (ALPs). 
This framework can be realized in the type IIB string axiverse with hierarchical axion masses and decay constants.
Our investigation reveals that in the light QCD axion scenario, the energy density of the lightest ALP often dominates after mixing.
On the other hand, in the heavy QCD axion scenario, both the QCD axion and non-lightest ALPs may dominate, depending on the ALP decay constants.
Under certain conditions, the QCD axion can dominate the DM budget. 
Finally, we briefly discuss a theoretical framework featuring $\sim\mathcal{O}(100)$ axions, with hierarchical axion masses and decay constants.
 		
\vspace{3.1cm}
			
\bigskip
\noindent\today
\end{titlepage}
			
\setcounter{tocdepth}{2}
			
 

\section{Introduction}

Evidence for dark matter (DM) has been accumulating for a long time, yet its true nature remains unknown. 
The lack of supersymmetric signals at the LHC has undermined weakly interacting massive particle (WIMP) as a prime candidate, broadening the range of possibilities, with the axion somewhat taking its place.
See ref.~\cite{Cirelli:2024ssz} for a recent review on DM.

The axion, often referred to as the QCD axion, was originally postulated through the Peccei-Quinn (PQ) mechanism, which was devised to dynamically resolve the strong CP problem in the Standard Model (SM) \cite{Peccei:1977hh, Peccei:1977ur, Weinberg:1977ma, Wilczek:1977pj}.
Furthermore, a substantial quantity of ultra-light axion-like particles (ALPs) can emerge from higher-dimensional gauge fields, with the QCD axion candidate naturally being among them \cite{Witten:1984dg, Green:1984sg, Choi:2003wr, Reece:2024wrn}.
As is widely known, string theory predicts the existence of a plenitude of axions \cite{Svrcek:2006yi, Conlon:2006tq}, collectively referred to as the string axiverse \cite{Arvanitaki:2009fg, Acharya:2010zx, Cicoli:2012sz}.

In recent years, the QCD axion and ALPs have garnered widespread attention as natural candidates for cold DM due to their non-thermal production in the early Universe via the misalignment mechanism \cite{Preskill:1982cy, Abbott:1982af, Dine:1982ah}.
In the early Universe, the oscillation of a coherent axion state within its potential well can significantly contribute to the overall DM density. 
The equation of motion (EOM) governing this state resembles that of a damped harmonic oscillator. 
Here, the damping term is directly proportional to the Hubble parameter, while the restoring force is proportional to the axion mass.
As the Universe cools down, once the Hubble parameter becomes comparable to the axion mass, the system transitions from an overdamped regime, where oscillations are heavily suppressed, to an underdamped regime, characterized by the onset of sustained oscillations. 
These oscillations are pivotal, as they provide a plausible explanation for the observed abundance of cold DM.
See $\rm e.g.$ refs.~\cite{Marsh:2015xka, OHare:2024nmr} for reviews.

For QCD axion DM, assuming an initial misalignment angle of order unity, the aforementioned misalignment mechanism imposes an upper limit on the classical QCD axion window, given by $f_a\lesssim 10^{11}-10^{12}\, {\rm GeV}$, where $f_a$ represents the QCD axion decay constant. 
In other words, one of the challenges for QCD axion DM is the issue of overproduction; this problem arises when considering an axion decay constant larger than $10^{12}\, {\rm GeV}$ \cite{OHare:2024nmr}. 
In addition, if we consider the recently proposed dark dimension scenario \cite{Montero:2022prj}, in which the QCD axion decay constant is constrained to a narrow range $f_a\sim 10^{9}-10^{10}\, \rm GeV$, then QCD axion DM will face the problem of insufficient abundance \cite{Gendler:2024gdo, Li:2024jko}.
However, both of these issues concerning the QCD axion DM can be naturally explained through axion mixing. 
In the light QCD axion scenario \cite{Daido:2015cba, Li:2023uvt}, the abundance of the QCD axion can be effectively suppressed, whereas in the heavy QCD axion scenario \cite{Cyncynates:2023esj}, its abundance can be significantly enhanced.
It should be noted that, in this context, axion mixing generally refers to the mixing between two axions, namely the mass mixing effect between the QCD axion and an ALP \cite{Hill:1988bu, Ho:2018qur, Li:2023xkn, Li:2023det, Li:2024psa, Murai:2024nsp, Murai:2025wbg}.
See also ref.~\cite{Li:2025cep} for a recent discussion on multi-axion mixing, which involves one QCD axion and multiple ultra-light ALPs. 

In this work, we introduce a mechanism for multi-component DM, which arises from multi-axion mixing, and examine some of its defining properties.
To begin with, we review the single-component axion DM scenario, involving the QCD axion and ALP, via the misalignment mechanism.
Subsequently, we shift our attention to the multi-component axion DM scenario.
Here multi-component DM implies that the cold DM is constituted by a combination of the QCD axion and many ultra-light ALPs. 
In the simplest two-component axion DM framework, the energy density of the mixed ALP dominates in the light QCD axion scenario, whereas the QCD axion dominates in the heavy QCD axion scenario.
Expanding our analysis to multi-component axion DM frameworks, we find that in the light QCD axion scenario, the energy density of the lightest ALP often dominates after mixing.
In contrast, in the heavy QCD axion scenario, both the QCD axion and non-lightest ALPs can potentially dominate the energy density, depending on the relationship between the decay constants of the ALPs.
Under certain conditions, the QCD axion can dominate the DM budget.
Finally, we briefly discuss a theoretical framework that incorporates $\sim\mathcal{O}(100)$ axions, with hierarchical axion masses and decay constants in the type IIB string axiverse.

The rest of this paper is structured as follows.  
In section~\ref{one_component}, we briefly review the single-component axion DM scenario.
In section~\ref{multi_component}, we delve into the multi-component scenario, including two-, three-, and multi-component
cases.
In section~\ref{sec_framework}, we briefly discuss a multi-axion framework with hierarchical axion masses and decay constants.
Finally, the conclusion is given in section~\ref{sec_Conclusion}.
In appendix~\ref{appendix_A}, as an example, we show more details of the three-component axion DM scenario, including the axion mixing potential and the mass mixing matrix.

\section{Single-component axion dark matter}
\label{one_component}

In this section, we briefly review the single-component axion DM scenario, encompassing both the QCD axion and ALP DM, via the misalignment mechanism. 
It is important to note that in this work, we solely focus on the pre-inflationary scenario in which the PQ symmetry is spontaneously broken during inflation.

According to the misalignment mechanism \cite{Preskill:1982cy, Abbott:1982af, Dine:1982ah}, in the single-component axion DM scenario, the present energy density of the QCD axion ($a_0$) and the ALP\footnote{Notice that the ALP discussed here is a single-field particle with a constant mass. For the sake of convenience in subsequent discussions, we denote it as $A_i$, and we can set $i=1$ in this section.} ($A_1$) can be respectively expressed as \cite{Li:2023xkn}
\begin{eqnarray}
\text{QCD axion:}~\rho_{a_0,0}&=&\dfrac{1}{2} m_{a_0,0} m_{a_0,{\rm osc}} f_{a_0}^2 \theta_{{\rm i},a_0}^2 \left(\dfrac{\mathfrak{a}_{{\rm osc},a_0}}{\mathfrak{a}_0}\right)^3\, ,\\
\text{ALP:}~\rho_{A_1,0}&=&\dfrac{1}{2} m_{A_1}^2 f_{A_1}^2 \Theta_{{\rm i},A_1}^2 \left(\dfrac{\mathfrak{a}_{{\rm osc},A_1}}{\mathfrak{a}_0}\right)^3\, ,
\end{eqnarray}
where $m_{a_0,0}$ and $m_{A_1}$ are the QCD axion zero-temperature mass and ALP mass, respectively, $f_{a_0}$ and $f_{A_1}$ are the axion decay constants, $\theta_{\rm i}$ and $\Theta_{\rm i}$ are the initial misalignment angles, $m_{a_0,{\rm osc}}$ represents the QCD axion mass at the oscillation temperature $T_{{\rm osc}, {a_0}}$, 
$\mathfrak{a}_{{\rm osc}}$ and $\mathfrak{a}_0$ are the scale factors at $T_{{\rm osc}}$ and at the current CMB temperature $T_0$.
Note that here we have not taken into account some numerical factors. 
For the QCD axion, in order to explain the observed DM abundance with $f_{a_0}=1\times10^{12}\, \rm GeV$, the initial misalignment angle should be $\sim\mathcal{O}(1)$. 
In the subsequent discussions, we will assume that the initial misalignment angles ($\theta_{\rm i}$ and $\Theta_{\rm i}$) are of order one for all cases.  
 
We find that in the single-component axion DM scenario, the ratio of the energy densities between the QCD axion and ALP is given by
\begin{eqnarray} 
\dfrac{\rho_{a_0,0}}{\rho_{A_1,0}}\simeq \dfrac{m_{a_0,0}}{\sqrt{m_{a_0,{\rm osc}} m_{A_1}}} \left(\dfrac{f_{a_0}}{f_{A_1}}\right)^2\, .
\end{eqnarray} 
Since we will later consider a model with hierarchical axion decay constants, we are more concerned about the two scenarios: either the QCD axion decay constant is significantly larger than that of the ALP, $\rm i.e.$, $f_{a_0}\gg f_{A_1}$, or it is significantly smaller, $\rm i.e.$, $f_{a_0}\ll f_{A_1}$.
In both scenarios, we find that different axions {\it dominate} in the single-component axion DM scenario: 
\begin{eqnarray}
f_{a_0}&\gg& f_{A_1}:~\text{QCD axion dominated},\\ 
f_{a_0}&\ll& f_{A_1}:~\text{ALP dominated}.
\end{eqnarray}
Notice that in this context, ``dominated$"$ merely indicates which scenario has a greater axion energy density between the two. 
It should not be confused with the subsequent multi-component axion DM scenarios, where ``dominated$"$ truly signifies which axion energy density is dominant. 
 
We close this section with a brief discussion of axion mixing in the single-component axion DM scenario.
Axion mixing has been extensively discussed within multi-axion frameworks, particularly in the scenario involving two axions --- one QCD axion and one ALP. 
In these frameworks, the axion decay constants are considered to be hierarchical, with $f_{a_0} \gg f_{A_1}$ corresponding to the light QCD axion scenario \cite{Daido:2015cba, Li:2023uvt}, and $f_{a_0} \ll f_{A_1}$ corresponding to the heavy QCD axion scenario \cite{Cyncynates:2023esj}. 
It is important to note that while the mixing of two axions is considered in that context, ultimately, only the abundance of a single axion, namely the QCD axion, is taken into account, whereas the abundance of the ALP is not considered. 
Therefore, we believe it is necessary to account for the energy densities of both, which is also the focus of this work. 
Additionally, we have recently investigated the mixing of multiple axions, where the number of axions exceeds two, typically consisting of one QCD axion and multiple ALPs \cite{Li:2025cep}. 
Hence, another key emphasis of this work is to study the axion energy densities in such multi-axion mixing scenarios. 
 
\section{Multi-component axion dark matter}
\label{multi_component}   
  
In this section, we delve into the multi-component axion DM scenario. 
Here, ``multi-component$"$ implies that the DM is composed of the QCD axion and ALPs. 
Generally, we can consider the scenario where one QCD axion coexists with one or more ALPs --- this is achievable in the string axiverse, especially in the multi-axion mixing case considered in ref.~\cite{Li:2025cep}. 
However, only the situation of QCD axion DM was briefly discussed. 
Therefore, this work is based on that paper but focuses on the multi-component DM scenario. 
In addition, the properties of QCD axion or ALP DM in the case of two-axion mixing have been discussed in previous literature, but all these discussions were based on a single-component axion DM scenario. 
Consequently, the nature of the multi-component axion DM we discuss here has not been explored in previous literature.  
  
Taking into account the considerations in ref.~\cite{Li:2025cep}, the low-energy effective potential for the multi-component axion DM model here can be expressed as
\begin{eqnarray}
V_{\rm eff}=\sum_i\Lambda_i^4\left[1-\cos\left(\sum_j n_{ij} \theta_j \right)\right]\, ,
\end{eqnarray} 
where $\theta_i$, $f_{\theta_i}$, $\Lambda_i$, and $n_{ij}$ represent the axion angles, axion decay constants, overall scales, and domain wall numbers, respectively.
Here, we consider the mixing scenario involving one QCD axion and $N$ ALPs. 
Therefore, the matrix of domain wall numbers $n_{ij}$ should be taken as a $(N+1)\times(N+1)$ matrix $\mathfrak{n}_{\mathfrak{ij}}$, beginning with indices $\mathfrak{i}=0$ and $\mathfrak{j}=0$.
More details on the mixing mechanism can be found in the original paper.
Here we will focus solely on the discussion of axion energy density.
In this multi-component axion DM scenario, the present total axion energy density can be expressed as
\begin{eqnarray}
\rho_{\rm tot,0}=\rho_{a_0,0}+\sum_i^N \rho_{A_i,0}\, .
\end{eqnarray} 
Notice that here, $\rho_{a_0,0}$ and $\rho_{A_i,0}$ represent the present axion energy densities after taking into account the axion mixing effect.

In addition, it should be emphasized that in this work, we assume that the transfer of energy density between axions during the mixing process is adiabatic. 
The adiabatic condition in the two-axion mixing process can be satisfied \cite{Ho:2018qur}, namely, $\Delta t_\times \gg \max [2\pi/ m_l(T_\times), \, 2\pi/(m_h(T_\times)-m_l(T_\times))]$, where $\Delta t_\times$ represents the duration of level crossing.
It implies that the comoving axion numbers of the eigenstates $a_h$ (with mass eigenvalue $m_h$) and $a_l$ (with $m_l$) are individually conserved at the level crossing temperature $T_\times$.
Furthermore, in the recent ref.~\cite{Murai:2024nsp}, the adiabatic condition in the context of level crossing between the QCD axion and ALP is further discussed. 
By clarifying the relationships among various bases for axion level crossing descriptions, they established an improved, basis-independent adiabatic condition definition that surpasses the limitations of previous formulations.

However, in our multi-component axion DM scenario, for simplicity, we consider either $f_{a_0} \gg f_{A_i}$ or $f_{a_0} \ll f_{A_i}$, that is, there exists a large hierarchy between the decay constants of the QCD axion and the ALPs.
In this case, the adiabatic condition will not be violated, implying that the energy transfer between axions is complete. 
This also holds true in the multi-axion mixing scenario. 
Regarding the extent to which the adiabatic condition is violated, this remains a subject for further numerical analysis. 
On the other hand, if the adiabatic condition is severely violated, that is, in the extreme case where there is no energy transfer between axions, then the multi-component axion DM scenario discussed here will no longer hold. 
Therefore, the assumption we make here is, ideally, that there is a large hierarchy in the axion decay constants, and this assumption can be naturally realized in the string axiverse.

Next, we will separately explore scenarios for two-, three-, and multi-component axion DM, taking into account both the light and heavy QCD axion frameworks.

\subsection{Two-component}  

We first consider the simplest two-component axion DM scenario --- involving one QCD axion ($a_0$) and one ALP ($A_1$).
In this case, the present energy density of the QCD axion and ALP after mixing can be respectively expressed as \cite{Li:2023xkn}
\begin{eqnarray}
\text{QCD axion:}~\rho_{a_0,0}&=&\dfrac{1}{2} m_{a_0,0} m_{A_1}  f_{A_1}^2 \Theta_{{\rm i},A_1}^2 \left(\dfrac{\mathfrak{a}_{{\rm osc},A_1}}{\mathfrak{a}_0}\right)^3\, ,\\
\text{ALP:}~\rho_{A_1,0}&=&\dfrac{1}{2} m_{A_1} m_{a_0,{\rm osc}} f_{a_0}^2 \theta_{{\rm i},a_0}^2 \left(\dfrac{\mathfrak{a}_{{\rm osc},a_0}}{\mathfrak{a}_0}\right)^3\, .
\label{two-component_ALP}
\end{eqnarray}
It is worth noting that the scenario in ref.~\cite{Li:2023xkn} differs slightly from what we have here. 
In this scenario, we are considering the simplest situation of two-axion mixing.
Then the ratio of the energy densities between the QCD axion and ALP is given by
\begin{eqnarray} 
\dfrac{\rho_{a_0,0}}{\rho_{A_1,0}}\simeq \dfrac{m_{a_0,0} \sqrt{m_{a_0,{\rm osc}}}}{m_{A_1}^{3/2}} \left(\dfrac{f_{A_1}}{f_{a_0}}\right)^2\, ,
\end{eqnarray} 
where the initial misalignment angles are assumed to be of order one.
We find that different axions dominate in the two-component axion DM scenario: 
\begin{eqnarray}
f_{a_0}&\gg& f_{A_1}:~\text{ALP dominated},\\ 
f_{a_0}&\ll& f_{A_1}:~\text{QCD axion dominated},
\end{eqnarray}
corresponding to the light and heavy QCD axion scenarios, respectively.
Notice that, for the effective mixing to occur, we need to assume that the mass of the ALP is slightly smaller than the zero-temperature mass of the QCD axion.

Here, we need to stress once again that although the two-axion mixing effect within the single-component axion DM scenario has been previously discussed, the situation changes when considering the multi-component scenario. 
The aforementioned equations reveal that in the light QCD axion scenario, when examining a two-component axion DM model, the QCD axion no longer dominates. 
Instead, it is the mixed ALP that becomes the dominant component of DM. 
However, we find that in the heavy QCD axion scenario, the mixed QCD axion still holds a dominant position.
 
\begin{figure}[t]
\centering
\includegraphics[width=0.70\textwidth]{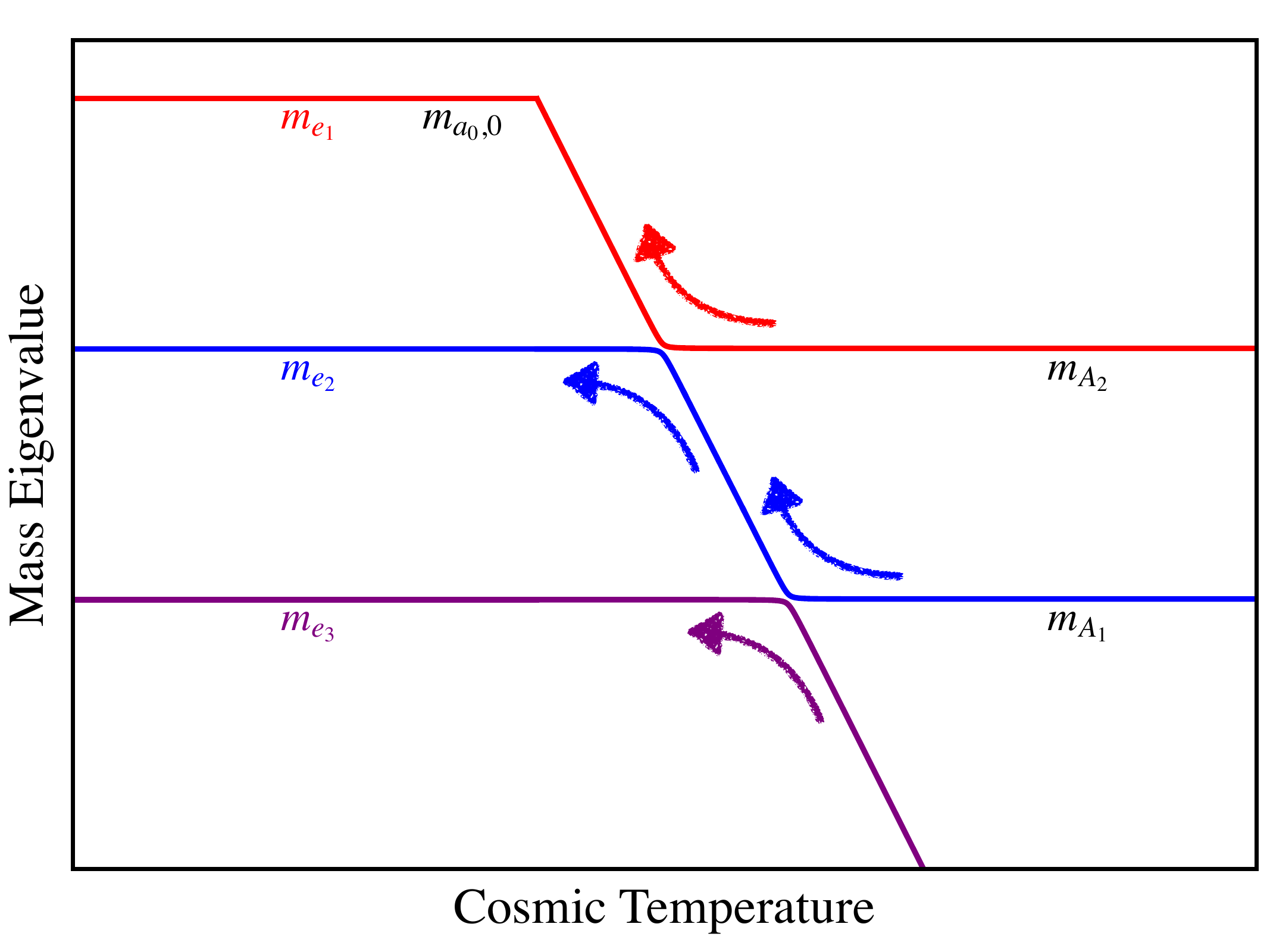}
\caption{Schematic illustration regarding the transfer of axion energy density in the three-component axion DM scenario.
Notice that the evolutionary processes in the light and heavy QCD axion scenarios are similar, with the main difference lying in the relationship between the axion decay constants.
The red, blue, and purple solid lines represent the mass eigenvalues $m_{e_i}$, and the corresponding arrows indicate the directions of axion energy density transfer.
The evolution direction of the cosmic temperature is from right to left.
This plot can also be extended to the multi-component scenario.}
\label{fig_three-component}
\end{figure} 
 
\subsection{Three-component}

In this subsection, we discuss the three-component axion DM scenario --- involving one QCD axion ($a_0$) and two ALPs ($A_1$ and $A_2$, where we assume $m_{A_1}<m_{A_2}<m_{a_0,0}$).
More details about this scenario can be found in appendix~\ref{appendix_A}.
See also figure~\ref{fig_three-component} for a schematic illustration regarding the transfer of axion energy density in this case.

It is necessary here to present the energy densities of axions separately after mixing.
We first discuss the QCD axion and need to start with the ALP $A_2$.
The initial energy density of $A_2$ at the oscillation temperature $T_{{\rm osc}, {A_2}}$ is expressed as
\begin{eqnarray}
\rho_{{A_2},{\rm osc}}=\frac{1}{2}m_{A_2}^2 f_{A_2}^2 \Theta_{{\rm i},A_2}^2\, ,
\end{eqnarray}
with $\Theta_{{\rm i},A_2}$ representing the initial misalignment angle of $A_2$.
Subsequently, in the temperature range $T_{\times_2}<T<T_{{\rm osc}, {A_2}}$, the energy density of $A_2$ remains adiabatically invariant.
At the temperature $T_{\times_2}$, where the level crossing occurs, this energy density can be characterized by
\begin{eqnarray}
\rho_{A_2,\times_2}=\frac{1}{2}m_{A_2}^2 f_{A_2}^2 \Theta_{{\rm i},A_2}^2 \left(\frac{\mathfrak{a}_{{\rm osc},A_2}}{\mathfrak{a}_{\times_2}}\right)^3 \, ,
\end{eqnarray}
where $\mathfrak{a}_{{\rm osc},A_2}$ and $\mathfrak{a}_{\times_2}$ denote the scale factors at $T_{{\rm osc},A_2}$ and $T_{\times_2}$, respectively.
Following this, the energy density $\rho_{A_2,\times_2}$ is transferred to the QCD axion. 
The axion energy density then stays adiabatically invariant up to the current CMB temperature $T_0$.
As a result, we arrive at the present QCD axion energy density
\begin{eqnarray}
\text{QCD axion:}&& \rho_{a_0,0}=\frac{1}{2}m_{a_0,0}m_{A_2} f_{A_2}^2 \Theta_{{\rm i},A_2}^2 \left(\frac{\mathfrak{a}_{{\rm osc},A_2}}{\mathfrak{a}_0}\right)^3 \, ,
\end{eqnarray}
where $\mathfrak{a}_0$ is the scale factor at $T_0$. 

Then we discuss the ALP $A_2$ and need to start with the ALP $A_1$.
The initial energy density of $A_1$ at the oscillation temperature $T_{{\rm osc}, {A_1}}$ is expressed as
\begin{eqnarray}
\rho_{{A_1},{\rm osc}}=\frac{1}{2}m_{A_1}^2 f_{A_1}^2 \Theta_{{\rm i},A_1}^2\, ,
\end{eqnarray}
with $\Theta_{{\rm i},A_1}$ representing the initial misalignment angle of $A_1$.
In the temperature range $T_{\times_1}<T<T_{{\rm osc}, {A_1}}$, the energy density of $A_1$ is adiabatically invariant, which at the level crossing temperature $T_{\times_1}$ is given by
\begin{eqnarray}
\rho_{A_1,\times_1}=\frac{1}{2}m_{A_1}^2 f_{A_1}^2 \Theta_{{\rm i},A_1}^2 \left(\frac{\mathfrak{a}_{{\rm osc},A_1}}{\mathfrak{a}_{\times_1}}\right)^3 \, .
\end{eqnarray}
Subsequently, the energy density $\rho_{A_1,\times_1}$ is transferred to the QCD axion and stays adiabatically invariant up to the temperature $T_{\times_2}$.
We can obtain the corresponding QCD axion energy density
\begin{eqnarray}
\rho_{a_0,\times_2}=\frac{1}{2}m_{A_1} m_{A_2} f_{A_1}^2 \Theta_{{\rm i},A_1}^2 \left(\frac{\mathfrak{a}_{{\rm osc},A_1}}{\mathfrak{a}_{\times_2}}\right)^3 \, .
\end{eqnarray}
After $T_{\times_2}$, the energy density $\rho_{A_0,\times_2}$ is transferred to the ALP $A_2$ and stays adiabatically invariant up to $T_0$.
Therefore, we arrive at the present $A_2$ energy density 
\begin{eqnarray}
\text{ALP $A_2$:}&& \rho_{A_2,0}=\frac{1}{2}m_{A_1} m_{A_2} f_{A_1}^2 \Theta_{{\rm i},A_1}^2 \left(\frac{\mathfrak{a}_{{\rm osc},A_1}}{\mathfrak{a}_0}\right)^3 \, .
\end{eqnarray}

Next we discuss the ALP $A_1$ and need to start with the QCD axion $a_0$.
The initial energy density of $a_0$ at the oscillation temperature $T_{{\rm osc}, {a_0}}$ is expressed as
\begin{eqnarray}
\rho_{{a_0},{\rm osc}}=\frac{1}{2}m_{a_0,{\rm osc}}^2 f_{a_0}^2 \theta_{{\rm i},a_0}^2\, .
\end{eqnarray}
In the temperature range $T_{\times_1}<T<T_{{\rm osc}, {a_0}}$, the energy density of $a_0$ is adiabatically invariant, which at the level crossing temperature $T_{\times_1}$ is given by
\begin{eqnarray}
\rho_{a_0,\times_1}=\dfrac{1}{2} m_{A_1} m_{a_0,{\rm osc}} f_{a_0}^2 \theta_{{\rm i},a_0}^2 \left(\dfrac{\mathfrak{a}_{{\rm osc},a_0}}{\mathfrak{a}_{\times_1}}\right)^3 \, .
\end{eqnarray}
Then the energy density $\rho_{a_0,\times_1}$ is transferred to the ALP $A_1$ and stays adiabatically invariant up to the temperature $T_0$.
Therefore, we arrive at the present $A_1$ energy density 
\begin{eqnarray}
\text{ALP $A_1$:}&& \rho_{A_1,0}=\dfrac{1}{2} m_{A_1} m_{a_0,{\rm osc}} f_{a_0}^2 \theta_{{\rm i},a_0}^2 \left(\dfrac{\mathfrak{a}_{{\rm osc},a_0}}{\mathfrak{a}_0}\right)^3 \, ,
\end{eqnarray}
which is analogous to the situation considered in the two-component axion DM scenario, as shown in eq.~\eqref{two-component_ALP}.

Now the ratios of the energy densities between the QCD axion and ALPs are respectively expressed as
\begin{eqnarray} 
\dfrac{\rho_{a_0,0}}{\rho_{A_2,0}}&\simeq& \dfrac{m_{a_0,0} \sqrt{m_{A_1}}}{m_{A_2}^{3/2}} \left(\dfrac{f_{A_2}}{f_{A_1}}\right)^2\, ,\\
\dfrac{\rho_{a_0,0}}{\rho_{A_1,0}}&\simeq& \dfrac{m_{a_0,0} \sqrt{m_{a_0,{\rm osc}}}}{m_{A_1} \sqrt{m_{A_2}}} \left(\dfrac{f_{A_2}}{f_{a_0}}\right)^2\, ,\\
\dfrac{\rho_{A_2,0}}{\rho_{A_1,0}}&\simeq& \dfrac{m_{A_2} \sqrt{m_{a_0,{\rm osc}}}}{m_{A_1}^{3/2}} \left(\dfrac{f_{A_1}}{f_{a_0}}\right)^2\, ,
\end{eqnarray} 
where the initial misalignment angles are assumed to be of order one.
We find that different axions dominate in the three-component axion DM scenario: 
\begin{eqnarray}
f_{a_0}&\gg& f_{A_1}\sim f_{A_2}:~\text{ALP $A_1$ dominated},\\ 
f_{a_0}&\ll& f_{A_1}\sim f_{A_2}:~\text{QCD axion and ALP $A_2$ dominated},
\end{eqnarray}
corresponding to the light and heavy QCD axion scenarios, respectively.
It should be noted that, according to ref.~\cite{Li:2025cep}, all the ALP decay constants here are either simultaneously smaller than the QCD axion decay constant or simultaneously larger than it.
We find that in the light QCD axion scenario, only the ALP $A_1$ is dominant. 
In contrast, in the heavy QCD axion scenario, there are two distinct cases depending on the relationship between the decay constants of the ALPs $A_1$ and $A_2$.
Specifically, when $f_{A_2}$ is significantly larger than $f_{A_1}$, the QCD axion is dominant. 
Conversely, when $f_{A_2}$ is significantly smaller than $f_{A_1}$, the ALP $A_2$ is dominant. 
This can be represented as follows:
\begin{eqnarray}
f_{a_0}&\ll& f_{A_1}\ll f_{A_2}:~\text{QCD axion dominated},\\
f_{a_0}&\ll& f_{A_2}\ll f_{A_1}:~\text{ALP $A_2$ dominated}.
\end{eqnarray}
Notice that in the light QCD axion scenario, regardless of whether $f_{A_2}$ is larger or smaller than $f_{A_1}$, the ALP $A_1$ remains dominant.

When comparing the two-component and three-component axion DM scenarios, in the light QCD axion framework, there is always only one ALP that dominates.  
In the heavy QCD axion framework, for the QCD axion to dominate in the three-component scenario, it is required that $f_{A_1}$ is significantly smaller than $f_{A_2}$.

\subsection{Multi-component}
 
In this subsection, we discuss the more general multi-component axion DM scenario --- involving one QCD axion ($a_0$) and many ALPs ($A_i$, where we assume $m_{A_i}<m_{A_{i+1}}<m_{a_0,0},\, \forall i$). 
See also ref.~\cite{Li:2025cep} for more details on axion mixing.

In this case, as previously derived, the present energy density of the QCD axion and ALPs after mixing can be respectively expressed as 
\begin{eqnarray}
\text{QCD axion:}~\rho_{a_0,0}&=&\dfrac{1}{2} m_{a_0,0} m_{A_N}  f_{A_N}^2 \Theta_{{\rm i},A_N}^2 \left(\dfrac{\mathfrak{a}_{{\rm osc},A_N}}{\mathfrak{a}_0}\right)^3\, ,\\
\text{ALP $A_N$:}~\rho_{A_N,0}&=&\frac{1}{2}m_{A_{N-1}} m_{A_N} f_{A_{N-1}}^2 \Theta_{{\rm i},A_{N-1}}^2 \left(\frac{\mathfrak{a}_{{\rm osc},A_{N-1}}}{\mathfrak{a}_0}\right)^3 \, ,\\
\text{ALP $A_{N-1}$:}~\rho_{A_{N-1},0}&=&\frac{1}{2}m_{A_{N-2}} m_{A_{N-1}} f_{A_{N-2}}^2 \Theta_{{\rm i},A_{N-2}}^2 \left(\frac{\mathfrak{a}_{{\rm osc},A_{N-2}}}{\mathfrak{a}_0}\right)^3 \, ,
\end{eqnarray} 
\begin{eqnarray}
&\vdots&\nonumber~~~~~~~~~~~~~~~~~~~~~~~~~~~~~~~~~~~~~~~~~~~~~~~~~\\
\text{ALP $A_2$:}~\rho_{A_2,0}&=&\frac{1}{2}m_{A_1} m_{A_2} f_{A_1}^2 \Theta_{{\rm i},A_1}^2 \left(\frac{\mathfrak{a}_{{\rm osc},A_1}}{\mathfrak{a}_0}\right)^3 \, ,\\
\text{ALP $A_1$:}~\rho_{A_1,0} &=&\dfrac{1}{2} m_{A_1} m_{a_0,{\rm osc}} f_{a_0}^2 \theta_{{\rm i},a_0}^2 \left(\dfrac{\mathfrak{a}_{{\rm osc},a_0}}{\mathfrak{a}_0}\right)^3\, ,
\end{eqnarray} 
where $m_{A_N}$, $f_{A_N}$, and $\Theta_{{\rm i},A_N}$ represent the mass, the decay constant, and the initial misalignment angle of the ALP $A_N$, respectively.
Then the ratios of the energy densities between the QCD axion and ALPs are given by  
\begin{eqnarray} 
\dfrac{\rho_{a_0,0}}{\rho_{A_N,0}}&\simeq& \dfrac{m_{a_0,0} \sqrt{m_{A_{N-1}}}}{m_{A_N}^{3/2}} \left(\dfrac{f_{A_N}}{f_{A_{N-1}}}\right)^2\, ,\\
\dfrac{\rho_{a_0,0}}{\rho_{A_{N-1},0}}&\simeq& \dfrac{m_{a_0,0} \sqrt{m_{A_{N-2}}}}{m_{A_{N-1}}\sqrt{m_{A_N}}} \left(\dfrac{f_{A_N}}{f_{A_{N-2}}}\right)^2\, ,\\
&\vdots&\nonumber\\
\dfrac{\rho_{a_0,0}}{\rho_{A_2,0}}&\simeq& \dfrac{m_{a_0,0} \sqrt{m_{A_1}}}{m_{A_2} \sqrt{m_{A_N}}} \left(\dfrac{f_{A_N}}{f_{A_1}}\right)^2\, ,\\
\dfrac{\rho_{a_0,0}}{\rho_{A_1,0}}&\simeq& \dfrac{m_{a_0,0} \sqrt{m_{a_0,{\rm osc}}}}{m_{A_1} \sqrt{m_{A_N}}} \left(\dfrac{f_{A_N}}{f_{a_0}}\right)^2\, ,\\
\dfrac{\rho_{A_N,0}}{\rho_{A_{N-1},0}}&\simeq& \dfrac{m_{A_N} \sqrt{m_{N-2}}}{m_{A_{N-1}}^{3/2}} \left(\dfrac{f_{A_{N-1}}}{f_{A_{N-2}}}\right)^2\, ,\\
&\vdots&\nonumber\\
\dfrac{\rho_{A_3,0}}{\rho_{A_2,0}}&\simeq& \dfrac{m_{A_3} \sqrt{m_{A_1}}}{m_{A_2}^{3/2}} \left(\dfrac{f_{A_2}}{f_{A_1}}\right)^2\, ,\\
\dfrac{\rho_{A_2,0}}{\rho_{A_1,0}}&\simeq& \dfrac{m_{A_2} \sqrt{m_{a_0,{\rm osc}}}}{m_{A_1}^{3/2}} \left(\dfrac{f_{A_1}}{f_{a_0}}\right)^2\, ,
\end{eqnarray}  
where the initial misalignment angles are still assumed to be of order one. 
We also find that different axions dominate in the multi-component axion DM scenario: 
\begin{eqnarray}
f_{a_0}&\gg& f_{A_1}\sim\cdots\sim f_{A_N}:~\text{Lightest ALP $A_1$ dominated},\\ 
f_{a_0}&\ll& f_{A_1}\sim\cdots\sim f_{A_N}:~\text{QCD axion and ALPs (except $A_1$) dominated},
\end{eqnarray}
corresponding to the light and heavy QCD axion scenarios, respectively. 
As we discussed before, in the light QCD axion scenario, only the {\it lightest} ALP $A_1$ is dominant. 
In the heavy QCD axion scenario, both the QCD axion and the remaining ALPs have the potential to be dominant, and this depends on the relationship among the ALP decay constants. 
If all the ALP decay constants are very close to each other, then the QCD axion and all the ALPs (except $A_1$) are dominant. 
If there are significant differences among the ALP decay constants, it will lead to different axions being dominant, and this requires an analysis based on specific model parameters.

\begin{figure}[t]
\centering
\includegraphics[width=0.70\textwidth]{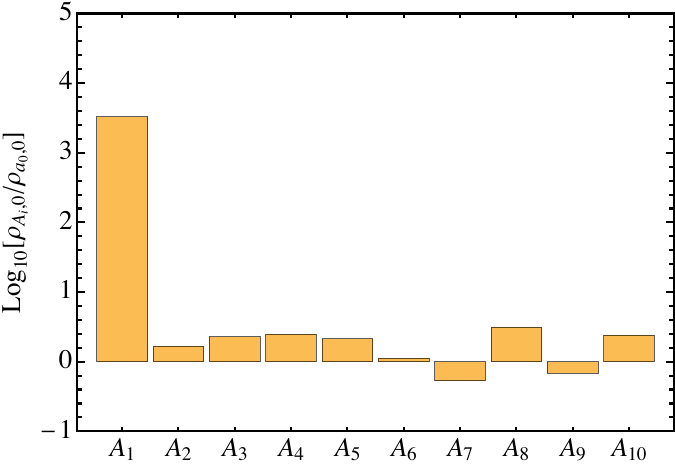}
\caption{The ratios of the energy densities $\rho_{A_i,0}/\rho_{a_0,0}$ in the light QCD axion scenario.
Here, we set $f_{a_0}=10^{12}\, {\rm GeV}$, and $f_{A_i}$ consist of 10 random numbers within the range of $10^{9.5}-10^{10.5}\, {\rm GeV}$.
Notice that we have neglected the contribution of axion masses in this context.}
\label{fig_ratio_1}
\end{figure}

\begin{figure}[h]
\centering
\includegraphics[width=0.70\textwidth]{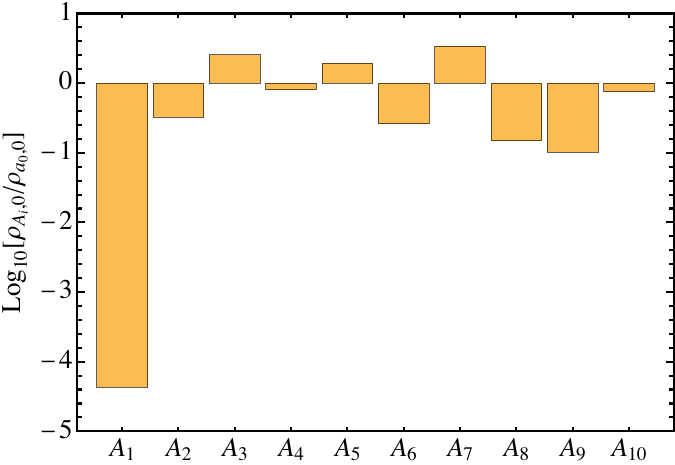}
\caption{Same as figure~\ref{fig_ratio_1} but for the heavy QCD axion scenario.
Here, we set $f_{a_0}=10^{12}\, {\rm GeV}$, and $f_{A_i}$ consist of 10 random numbers within the range of $10^{13.5}-10^{14.5}\, {\rm GeV}$.}
\label{fig_ratio_2}
\end{figure}  

In figures~\ref{fig_ratio_1} and \ref{fig_ratio_2}, we present schematic diagrams of the energy density ratio distributions in the light and heavy QCD axion scenarios, respectively.
Here, we consider the mixing scenario involving one QCD axion and ten ALPs.
In general, within the light QCD axion scenario, the energy density of the lightest ALP tends to dominate after mixing. 
In contrast, in the heavy QCD axion scenario, both the QCD axion and ALPs (non-lightest ones) can potentially dominate, depending on the relationship between the decay constants of the ALPs. 
Under certain conditions, the QCD axion can dominate the DM budget.

\section{A framework with hierarchical axion masses and decay constants} 
\label{sec_framework}

In this section, we briefly discuss a multi-axion framework with hierarchical axion masses and decay constants in the type IIB string axiverse.

The explicit multi-axion models within the type IIB string axiverse are discussed in detail in ref.~\cite{Cicoli:2012sz}, covering scenarios involving one QCD axion and one to two light ALPs. 
Furthermore, in a recent ref.~\cite{Broeckel:2021dpz}, we observe that multi-axion models with approximately $\mathcal{O}(100)$ axions are also achievable. 
In this case, the axion masses and decay constants exhibit a logarithmic distribution rather than a power-law one.
Specifically, the geometric requirement is the existence of two blow-up modes: one ($D_{\rm SM}$) to accommodate the SM and another ($D_{\rm np}$) to support non-perturbative effects.
One can consider the internal volume
\begin{eqnarray}
\mathcal{V}=\dfrac{1}{6}\sum_{i,j,k=1}^{h^{1,1}-2} k_{ijk} t_i t_j t_k-\gamma_{\rm SM} \tau_{\rm SM}^{3/2}-\gamma_{\rm np}\tau_{\rm np}^{3/2}\, ,
\end{eqnarray} 
where $h^{1,1}$ is the number of Kähler moduli, $k_{ijk}$ are the intersection numbers, $t_i$ are 2-cycle volumes, $\tau_i$ are 4-cycle moduli, and $\gamma_i$ are coefficients.
The leading contributions to the scalar potential arise from the corrections to the Kähler potential $K$ and the superpotential $W$, which stabilize three moduli $\mathcal{V}$, $\tau_{\rm np}$, and $\theta_{\rm np}$.

Since the axion $\theta_{\rm np}$ is too heavy, we need to consider stabilizing the remaining Kähler moduli by incorporating additional subleading contributions to $K$ and $W$.
The total potential can be expressed as \cite{Broeckel:2021dpz}
\begin{eqnarray}
V_{\rm tot}=V_{\rm LVS}(\mathcal{V})+V_F(\mathcal{V},t_i)\, ,
\end{eqnarray} 
where $V_{\rm LVS}$ represents the LARGE volume scenario (LVS) potential, and $V_F$ represents the F-term contribution.
The resulting axiverse features a perfect QCD axion $\theta_{\rm SM}$ along with multiple ultra-light ALPs, and their masses are respectively given by
\begin{eqnarray}
\text{QCD axion:}~m_{a_0,0}&\simeq&\dfrac{\Lambda_{\rm QCD}^2}{f_{a_0}}\, ,\\
\text{ALPs:}~m_{A_i}&\simeq&M_p e^{-\pi \tau_i/{\mathfrak{n}}_i}\, ,
\end{eqnarray} 
where $\Lambda_{\rm QCD}$ represents the scale of QCD confinement effects that yield the axion potential, $M_p$ is the Planck mass, and ${\mathfrak{n}}_i$ determines the periodicity of the cosine potential featuring a discrete shift symmetry.
The corresponding axion decay constants are respectively given by
\begin{eqnarray}
\text{QCD axion:}~f_{a_0}&=&\dfrac{c_{\rm SM}}{\tau_{\rm SM}^{1/4}}\dfrac{M_p}{\mathcal{V}^{1/2}}\, ,\\
\text{ALPs:}~f_{A_i}&=&\dfrac{c_i}{h_i}\dfrac{M_p}{\mathcal{V}^{2/3}}\, ,
\end{eqnarray} 
where $c_{\rm SM}$ and $c_i$ are coefficients of order unity, and $h_i$ are functions of the intersection numbers and the topological quantities.
When considering a scenario where the number of axions is of order $\sim\mathcal{O}(100)$, we should focus only on the region where $\mathcal{V}\gtrsim\mathcal{O}(10^{14})$, thereby obtaining a perfect QCD axion decay constant at the TeV-scale.

Notice that in this framework, the ALPs are ultra-light, whose masses are all smaller than that of the QCD axion, which is what we anticipate in our multi-component axion DM scenario.
Regarding the axion decay constants, we expect that some ALPs have decay constants smaller than that of the QCD axion, while others have decay constants larger than that of the QCD axion.
This corresponds respectively to the light QCD axion and heavy QCD axion scenarios discussed in the preceding sections. 
 
\section{Conclusion}
\label{sec_Conclusion}

In summary, we have introduced a mechanism for multi-component DM, which arises from multi-axion mixing, and examined some of its defining properties.
In this context, the multi-component axion DM scenario presents a fascinating departure from the single-component paradigm, as it posits that cold DM is composed of a combination of the QCD axion and numerous ultra-light ALPs. 
Through our analysis of the simplest two-component to multi-component frameworks, we uncovered significant implications in the light and heavy QCD axion scenarios. 
In the light QCD axion scenario, the mixing of axions often leads to the dominance of the energy density of the lightest ALP.
On the other hand, the heavy QCD axion scenario exhibits a more nuanced picture. 
Here, both the QCD axion and non-lightest ALPs have the potential to dominate the energy density, depending on the relationship between the decay constants of the ALPs. 
Under specific conditions, the QCD axion could be the main contributor to the DM amount.
Finally, our brief discussion of a theoretical framework that incorporates approximately $\mathcal{O}(100)$ axions within the type IIB string axiverse provides a potential foundation for our multi-component axion DM models.
The hierarchical axion masses and decay constants in this framework offer a rich playground for studying the dynamics of multi-axion systems and their implications for cold DM.
 
\section*{Acknowledgments}

This work was partly supported by the Institute of Theoretical Physics, CAS, and partly supported by the International Centre for Theoretical Physics Asia-Pacific.

\appendix

\section{More details about the three-component axion DM scenario}
\label{appendix_A}

Within this appendix, taking the three-component axion DM scenario as an example, we show more details about the axion mixing potential and the mass mixing matrix.
Consider one QCD axion ($a_0$) and two ALPs ($A_1$ and $A_2$).
In the light QCD axion scenario, the axion mixing potential is given by 
\begin{eqnarray}
\begin{aligned}
V_{\rm mix}&=m_{a_0}^2 f_{a_0}^2\left[1-\cos\left(\dfrac{\phi_0}{f_{a_0}}\right)\right]+m_{A_1}^2 f_{A_1}^2\left[1-\cos\left(\dfrac{\phi_0}{f_{a_0}}+\dfrac{\varphi_1}{f_{A_1}}\right)\right]\\
&+m_{A_2}^2 f_{A_2}^2\left[1-\cos\left(\dfrac{\phi_0}{f_{a_0}}+\dfrac{\varphi_2}{f_{A_2}}\right)\right]\, ,
\end{aligned}
\end{eqnarray} 
where $\phi_0$, $\varphi_1$, and $\varphi_2$ are the corresponding axion fields.
Expanding the potential to quadratic order gives the mass mixing matrix
\begin{eqnarray}
\mathbf{M}^2=
\left(\begin{array}{ccc}
\dfrac{m_{a_0}^2 f_{a_0}^2+m_{A_1}^2 f_{A_1}^2+m_{A_2}^2 f_{A_2}^2}{f_{a_0}^2} & ~ \dfrac{m_{A_1}^2 f_{A_1}}{f_{a_0}} & ~\dfrac{m_{A_2}^2 f_{A_2}}{f_{a_0}}\\
\dfrac{m_{A_1}^2 f_{A_1}}{f_{a_0}}   & ~m_{A_1}^2 & ~ 0\\
\dfrac{m_{A_2}^2 f_{A_2}}{f_{a_0}}   & ~ 0 & ~m_{A_2}^2
\end{array}\right)\, .
\label{mixing_matrix_light}
\end{eqnarray}
By considering the mixing of the QCD axion with each ALP separately, the mass mixing matrix can be expressed as two effective mixing matrices.
Subsequently, we can obtain the corresponding approximate mass eigenvalues $m_{e_i}$. 
It should be noted that a large hierarchy between the decay constants of the QCD axion and the ALPs needs to be taken into account here.
Similarly, in the heavy QCD axion scenario, the axion mixing potential is given by 
\begin{eqnarray}
\begin{aligned}
V_{\rm mix}&=m_{a_0}^2 f_{a_0}^2\left[1-\cos\left(\dfrac{\phi_0}{f_{a_0}}+\dfrac{\varphi_1}{f_{A_1}}+\dfrac{\varphi_2}{f_{A_2}}\right)\right]+m_{A_1}^2 f_{A_1}^2\left[1-\cos\left(\dfrac{\varphi_1}{f_{A_1}}\right)\right]\\
&+m_{A_2}^2 f_{A_2}^2\left[1-\cos\left(\dfrac{\varphi_2}{f_{A_2}}\right)\right]\, .
\end{aligned}
\end{eqnarray}  
Expanding the potential to quadratic order gives the mass mixing matrix 
\begin{eqnarray}
\mathbf{M}^2=
\left(\begin{array}{ccc}
m_{a_0}^2 & ~ \dfrac{m_{a_0}^2 f_{a_0}}{f_{A_1}} & ~\dfrac{m_{a_0}^2 f_{a_0}}{f_{A_2}}\\
\dfrac{m_{a_0}^2 f_{a_0}}{f_{A_1}} & ~\dfrac{m_{a_0}^2 f_{a_0}^2+m_{A_1}^2 f_{A_1}^2}{f_{A_1}^2} & ~\dfrac{m_{a_0}^2 f_{a_0}^2}{f_{A_1} f_{A_2}}\\
\dfrac{m_{a_0}^2 f_{a_0}}{f_{A_2}}   & ~ \dfrac{m_{a_0}^2 f_{a_0}^2}{f_{A_1} f_{A_2}} & ~\dfrac{m_{a_0}^2 f_{a_0}^2+m_{A_2}^2 f_{A_2}^2}{f_{A_2}^2}
\end{array}\right)\, .
\label{mixing_matrix_heavy}
\end{eqnarray}
We can observe that the evolutionary processes of axions in the light and heavy QCD axion scenarios are similar. 
However, due to the fundamental differences in the relationship as defined between the axion decay constants, the resulting phenomenological results are distinct.

\bibliographystyle{JHEP}
\bibliography{references}

\end{document}